\documentclass[twocolumn,twoside,slac_two]{revtex4}
\usepackage{graphicx}
\usepackage{fancyhdr}
\begin{document}

\title{Observations of Gamma-ray Bursts in the \textit{Fermi} era}


\author{G.Vianello}
\affiliation{Stanford University (HEPL), 452 Lomita Mall, Stanford, CA, 94205, US}
\author{on behalf of the \textit{Fermi}/LAT Collaboration}
\noaffiliation
\bigskip

\begin{abstract}
The \textit{Fermi} observatory, with its Gamma-Ray Bursts monitor (GBM) and Large Area Telescope (LAT), is observing Gamma-ray Bursts with unprecedented spectral coverage and sensitivity, from $\sim 10$ keV to $> 300$ GeV. In the first 3 years of the mission it observed emission above 100 MeV from 35 GRBs, an order of magnitude gain with respect to previous observations in this energy range. In this paper we review the main results obtained on such sample, highlighting also the relationships with the low-energy features (as measured by the GBM), and with measurements from observatories at other wavelengths. We also briefly discuss prospects for detection of GRBs by future Very-High Energy observatories such as HAWC and CTA, and by Gravitational Wave experiments.
\end{abstract}

\maketitle

\thispagestyle{fancy}

% References should be done using the \citep, \ref, and \label commands

\section{The \textit{Fermi} observatory}
\textit{Fermi} was launched on June 2008. It features two instruments: the Gamma-ray Burst Monitor (GBM) \citep{2009ApJ...702..791M}, a full sky monitor comprised of 12 sodium iodide (NaI) detectors and two bismuth germanate (BGO) detectors, sensitive respectively in the 8 keV - 1 MeV and 150 keV - 40 MeV energy range; and the Large Area Telescope (LAT) \citep{2009ApJ...697.1071A}, a pair production $\gamma$--ray telescope sensitive from 20 MeV to $> 300$ GeV. When compared to its precursor (the EGRET experiment onboard the Compton Gamma-Ray Observatory) the LAT features a larger field of view (2.4 sr at 1 GeV), a broader energy range, a lower dead time per event (27 $\mu$s) and a much larger (10x) effective area at all energy. As we will illustrate in the next section, this results in a much larger number of detections ($\sim$ 9 GRBs/year) with respect to EGRET (5 GRBs in 10 years).

\section{The \textit{Fermi}/LAT GRB catalog}
The first \textit{Fermi}/LAT GRB catalog \citep{2013arXiv1303.2908F} covers 3 years of observation, from August 2008 to July 2011. In such time period \textit{Fermi}/GBM detected $\sim$ 750 GRBs, with around half of them contained in the LAT field of view.  We used two detection algorithms: a standard likelihood algorithm, providing both detection and localization with $< 1$ deg accuracy, using Pass 6 v3 Transient events above 100 MeV; and a counting analysis using the LAT Low Energy (LLE) class of data, featuring a large effective area starting at $\sim$20 MeV but no localization capability. With the former analysis we detected and localized 28 GRBs, while using the latter analysis we detected 7 more bursts, for a total of 35 GRBs.
\subsection{High-energy emission}
While the number of GRB detected at high-energy by \textit{Fermi}/LAT is a small fraction of the total number of GRBs in the field of view, this sample allows us to uncover unique features of GRBs emerging only at high energies which we will summarize in the following sections.

\subsubsection{Energetics}
\begin{figure*}[tb!]
\includegraphics[width=\textwidth]{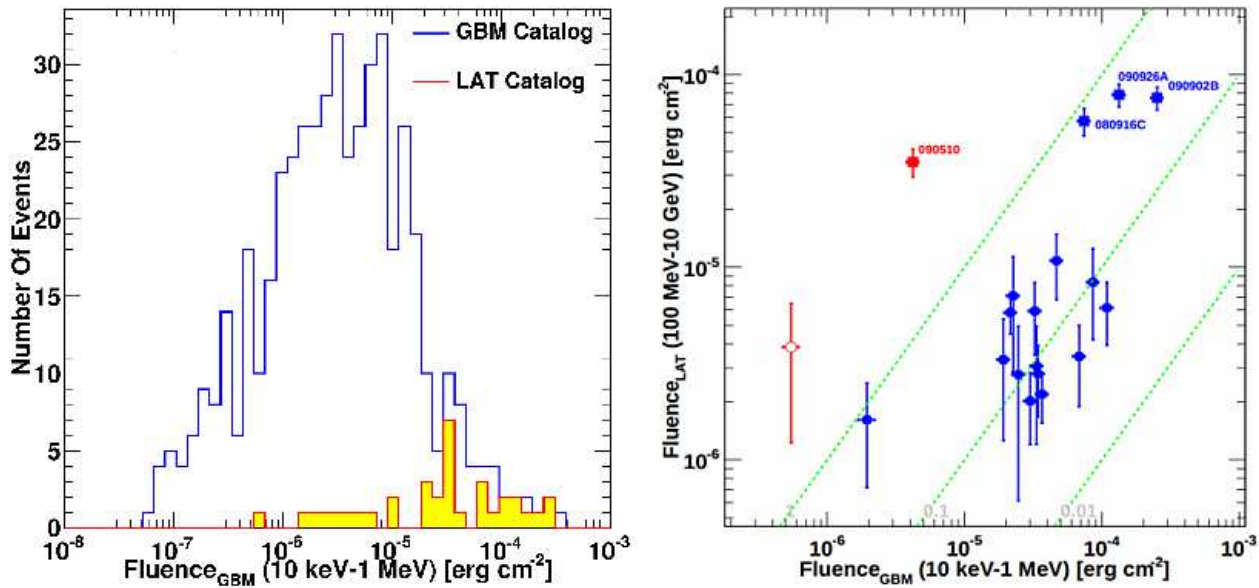}
\caption{Left panel: distribution of the low-energy fluence (10 keV - 1 MeV) for LAT detected GRBs (yellow), compared with the whole sample of GBM-detected GRBs \citep{2012ApJS..199...19G}. Right panel: high-energy fluence Vs low-energy fluence as observed by respectively LAT and GBM, for LAT-detected GRBs.}
\label{fig2}
\end{figure*}

Since LAT observations are photon-limited rather than background limited, the detection efficiency is directly related to the counts fluence of the source. This is an important difference with respect to \textit{Fermi}/GBM, which is background limited and for which the peak flux of the source is more relevant. Of course, the low-energy fluence is highly correlated with the high-energy fluence. The fact that the LAT detects preferentially GRBs with a high low-energy fluence (see left panel in Fig.~\ref{fig2}) is therefore not surprising. As shown in the right panel in Fig.~\ref{fig2}, the typical ratio between the high-energy fluence (above 100 MeV) and the low energy fluence (10 keV - 1 MeV) is $\sim 0.1$. It is interesting to note that there are four \textit{hyper-energetic} bursts for which the ratio exceed greatly the typical value, being closer or even above 1. These are GRB 080916C, 090510, 090902B and 090926A. The same conclusion can be reached taking the ratio of the rest frame total energy $E_{iso}$ in the two energy bands, which demonstrates that this is not an effect of the distance of these bursts, which are distributed between redshift 0.9 and 4.35. 

\subsubsection{Delayed and temporally extended emission}
\begin{figure*}[tb!]
\includegraphics[width=\textwidth]{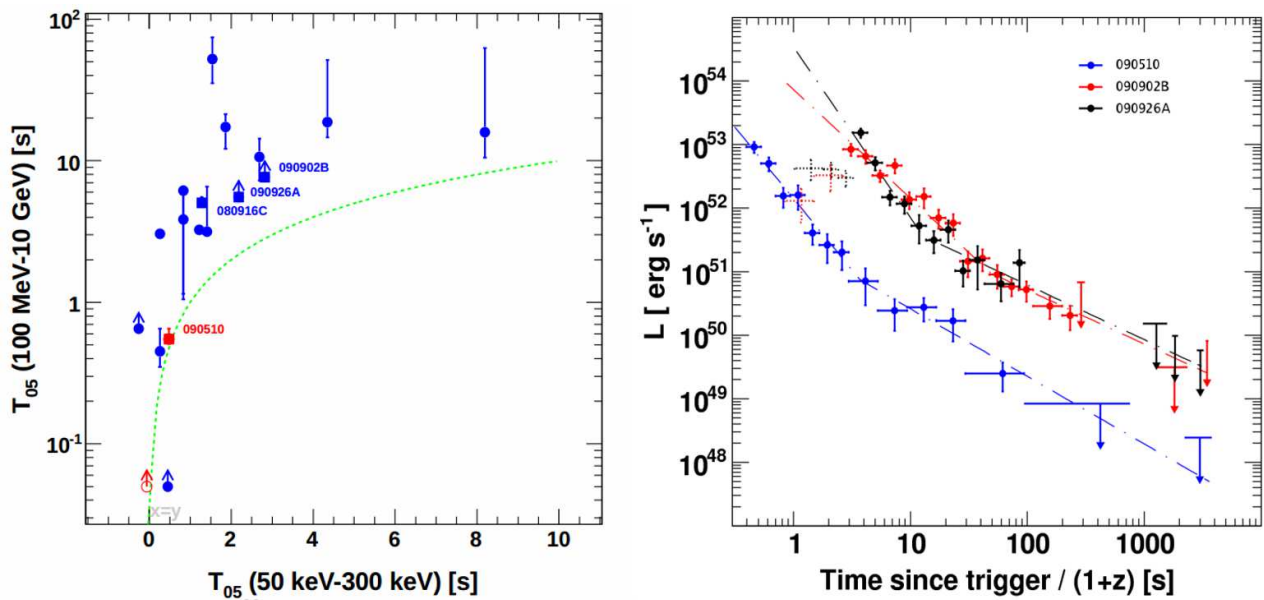}
\caption{Left panel: the onset of the high-energy emission (y-axis), as measured by T$_{05}$, appears to be sistematically delayed with respect to the low-energy onset (x-axis). The green line corresponds to x=y. Right panel: Luminosity as function of rest frame time for the three GRBs for which a broken power law (dashed lines) is a significantly better fit than a simple power law decay. }
\label{fig1}
\end{figure*}

The emission above 100 MeV is systematically delayed with respect to the low-energy emission. This can be seen in the left panel of Fig.~\ref{fig1}, where we used $T_{05}$ as a measure of the onset of the emission for both the 10-300 keV energy range (from \citet{2012ApJS..199...19G}) and the 100 MeV - 10 GeV energy range: it is clear that the latter is sistematically larger than the former.  Also, the duration of the high-energy emission appears to be sistematically longer, and features a smooth decaying phase after the end of the low-energy prompt emission. Such decaying phase is well described by a power law in all but three cases, for which we found that a broken power law describes better the data, as shown in the right panel of Fig.~\ref{fig1}. The significance of the breaks correspond to a chance probability of less than $10^{-5}$. Note that the time of the break is in all three cases after the end of the low-energy emission, as measured by T$_{90}$. If we define a \textit{late time decay index} $\alpha_{L}$ as the index of the power law for the light curves well described by a simple power law, and the index after the break for the three GRBs described by a broken power law, we find that $\alpha_{L} \sim -1$. This value is foreseen by the standard afterglow model for an adiabatic expansion of the fireball, while a radiative expansion would foreseen a decay with an index of 10/7, which is not observed in our data.

\subsubsection{High-energy photons}
The LAT has observed photons up to 30 GeV coming from bright GRBs, which in the case of high-redshift GRBs can become more than 100 GeV in the rest frame of the progenitor of the burst. This result poses a big challenge for the efficiency of the particle acceleration mechanisms, especially when considering the fact that some of this high-energy events have been detected within seconds since the start of the low-energy emission. In the context of the standard Fireball model \citep{1999PhR...314..575P} the presence of such high-energy photons constrains also the bulk Lorentz factor of the emitting shells to be $\Gamma > 1000$ in some cases, a value much higher than what previously thought. High energy photons from high redshift GRBs allow also to constrain the opacity of the Universe connected with the interaction of the $> 10$ GeV $\gamma$--rays with optical and UV photons of the Extragalactic Background Light (EBL). In the case of the short GRB 090510, the short time delay observed between low and high-energy events can be used to place tight limits on the energy dependence of the speed of light, which is postulated for example by some quantum gravity theories.

\subsection{Broadband spectroscopy}
\textit{Fermi} is an exceptional observatory for GRB spectroscopy. In particular, it has unprecedented spectral coverage, starting around 10 keV up to 300 GeV. We exploited this feature by performing a broad band spectral analysis of all the GRBs contained in the sample.
\subsubsection{Band model crisis}
Before \textit{Fermi} most of the GRB spectra were well described by the phenomenological Band model \citep{1993ApJ...413..281B}, which has become the \textit{de-facto} standard model. The spectra of all of the brightest bursts inside the LAT FoV present, on the contrary, signiﬁcant deviations from a Band function, requiring additional components such as power laws, high-energy cutoffs, or both. Other GRBs, observed at low off-axis angles, and with a corresponding high eﬀective area, show deviations as well. We conclude that the phenomenological Band model seems to be not sufficient to describe all the spectral features of LAT GRBs. Unfortunately, there is no common recipe, and different components can be required depending on the particular event. This calls for a better broad band modeling of the spectra of GRBs, opening new questions and prompting new theoretical developments.

\section{The afterglow of LAT-detected GRBs}
A subsample of LAT-detected GRBs have been studied at other wavelengths, in particular during their afterglow emission. A systematic study published by \citep{2011ApJ...738..138R} shown that in many ways the properties of the afterglow of LAT bursts are typical of the general afterglow population, but the ratio between the luminosity of the prompt emission and the luminosity of the afterglow is larger. Therefore, either their prompt emission is more efﬁcient in producing $\gamma$-rays, or, conversely, their afterglows are somehow soppressed. 
In two cases, GRB 090510 and GRB 110731, Swift and other instruments observed the afterglow when the high-energy extended emission was still detectable by the LAT. A broadband study, from optical wavelengths to $\gamma$-rays, showed that the emission is compatible with being from external shock \citep{2010ApJ...709L.146D,2013ApJ...763...71A}. In one other case, GRB 100728A, high-energy emission was detected by the LAT only in correspondence with an X--ray flare, which was successfully modeled from X--ray to $\gamma$-ray energies as internal shocks emission \citep{2011ApJ...734L..27A}.

\section{Multi-messenger observations, and synergies with other observatories}
Swift is routinely observing GRBs localized by \textit{Fermi}/LAT. Upon X-ray detection, the position is improved from a typical error of $0.3$ deg to sub-arcmin errors, which  consequently allows for measurement at other wavelengths (optical, IR...) and the measurement of the redshift for 1/4 of the LAT sample. As illustrated in the previous section, such observations can also be used to determine the origin of the extended emission observed by the LAT. In the near future, with the advent of new Very High-Energy (VHE) observatories such as HAWC and CTA, the energy coverage will be extended even more. The LAT large field of view, the presence of extra power-law components and very high-energy events, united with the large effective area of the new observatories will give good opportunities for a detection of GRBs with VHE observatories \citep{2012APh....35..641A,2012MNRAS.425..514K}. There are already a few examples of joint LAT-VHE observations, but no detections. 

\subsection{Gravitational waves}
Close-by short GRBs are thought to originate from the merger of two compact objects, such as Neutron Starts and Black Holes. Therefore, they are good candidates for the emission of detectable gravitational wave (GW) signals. The high effective area, duty cicle and detection efficiency of \textit{Fermi}/GBM, united with the localizing power of \textit{Fermi}/LAT, will be key elements in successfully detecting a GW signal from a short GRBs. Detectors currently in development, such as a-LIGO and \textit{advanced}-VIRGO, will be in the position of detecting such a signal in the very near future \citep{2012ApJ...760...12A,2013arXiv1304.0670L}.

\bigskip % extra skip inserted
\begin{acknowledgments}
The \textit{Fermi} LAT Collaboration acknowledges generous ongoing support
from a number of agencies and institutes that have supported both the
development and the operation of the LAT as well as scientific data analysis.
These include the National Aeronautics and Space Administration and the
Department of Energy in the United States, the Commissariat \`a l'Energie Atomique
and the Centre National de la Recherche Scientifique / Institut National de Physique
Nucl\'eaire et de Physique des Particules in France, the Agenzia Spaziale Italiana
and the Istituto Nazionale di Fisica Nucleare in Italy, the Ministry of Education,
Culture, Sports, Science and Technology (MEXT), High Energy Accelerator Research
Organization (KEK) and Japan Aerospace Exploration Agency (JAXA) in Japan, and
the K.~A.~Wallenberg Foundation, the Swedish Research Council and the
Swedish National Space Board in Sweden.

Additional support for science analysis during the operations phase is gratefully
acknowledged from the Istituto Nazionale di Astrofisica in Italy and the Centre National d'\'Etudes Spatiales in France.

\end{acknowledgments}

\bigskip % extra skip inserted
\bibliography{bib}

\begin{thebibliography}{14}
\expandafter\ifx\csname natexlab\endcsname\relax\def\natexlab#1{#1}\fi
\expandafter\ifx\csname bibnamefont\endcsname\relax
  \def\bibnamefont#1{#1}\fi
\expandafter\ifx\csname bibfnamefont\endcsname\relax
  \def\bibfnamefont#1{#1}\fi
\expandafter\ifx\csname citenamefont\endcsname\relax
  \def\citenamefont#1{#1}\fi
\expandafter\ifx\csname url\endcsname\relax
  \def\url#1{\texttt{#1}}\fi
\expandafter\ifx\csname urlprefix\endcsname\relax\def\urlprefix{URL }\fi
\providecommand{\bibinfo}[2]{#2}
\providecommand{\eprint}[2][]{\url{#2}}

\bibitem[{\citenamefont{{Meegan} et~al.}(2009)\citenamefont{{Meegan}, {Lichti},
  {Bhat}, {Bissaldi}, {Briggs}, {Connaughton}, {Diehl}, {Fishman}, {Greiner},
  {Hoover} et~al.}}]{2009ApJ...702..791M}
\bibinfo{author}{\bibfnamefont{C.}~\bibnamefont{{Meegan}}},
  \bibinfo{author}{\bibfnamefont{G.}~\bibnamefont{{Lichti}}},
  \bibinfo{author}{\bibfnamefont{P.~N.} \bibnamefont{{Bhat}}},
  \bibinfo{author}{\bibfnamefont{E.}~\bibnamefont{{Bissaldi}}},
  \bibinfo{author}{\bibfnamefont{M.~S.} \bibnamefont{{Briggs}}},
  \bibinfo{author}{\bibfnamefont{V.}~\bibnamefont{{Connaughton}}},
  \bibinfo{author}{\bibfnamefont{R.}~\bibnamefont{{Diehl}}},
  \bibinfo{author}{\bibfnamefont{G.}~\bibnamefont{{Fishman}}},
  \bibinfo{author}{\bibfnamefont{J.}~\bibnamefont{{Greiner}}},
  \bibinfo{author}{\bibfnamefont{A.~S.} \bibnamefont{{Hoover}}},
  \bibnamefont{et~al.}, \bibinfo{journal}{"Astrophys. J."}
  \textbf{\bibinfo{volume}{702}}, \bibinfo{pages}{791} (\bibinfo{year}{2009}),
  \eprint{0908.0450}.

\bibitem[{\citenamefont{{Atwood} et~al.}(2009)\citenamefont{{Atwood}, {Abdo},
  {Ackermann}, {Althouse}, {Anderson}, {Axelsson}, {Baldini}, {Ballet}, {Band},
  {Barbiellini} et~al.}}]{2009ApJ...697.1071A}
\bibinfo{author}{\bibfnamefont{W.~B.} \bibnamefont{{Atwood}}},
  \bibinfo{author}{\bibfnamefont{A.~A.} \bibnamefont{{Abdo}}},
  \bibinfo{author}{\bibfnamefont{M.}~\bibnamefont{{Ackermann}}},
  \bibinfo{author}{\bibfnamefont{W.}~\bibnamefont{{Althouse}}},
  \bibinfo{author}{\bibfnamefont{B.}~\bibnamefont{{Anderson}}},
  \bibinfo{author}{\bibfnamefont{M.}~\bibnamefont{{Axelsson}}},
  \bibinfo{author}{\bibfnamefont{L.}~\bibnamefont{{Baldini}}},
  \bibinfo{author}{\bibfnamefont{J.}~\bibnamefont{{Ballet}}},
  \bibinfo{author}{\bibfnamefont{D.~L.} \bibnamefont{{Band}}},
  \bibinfo{author}{\bibfnamefont{G.}~\bibnamefont{{Barbiellini}}},
  \bibnamefont{et~al.}, \bibinfo{journal}{"Astrophys. J."}
  \textbf{\bibinfo{volume}{697}}, \bibinfo{pages}{1071} (\bibinfo{year}{2009}),
  \eprint{0902.1089}.

\bibitem[{\citenamefont{{Fermi-LAT Collaboration}}(2013)}]{2013arXiv1303.2908F}
\bibinfo{author}{\bibnamefont{{Fermi-LAT Collaboration}}},
  \bibinfo{journal}{ArXiv e-prints}  (\bibinfo{year}{2013}),
  \eprint{1303.2908}.

\bibitem[{\citenamefont{{Goldstein} et~al.}(2012)\citenamefont{{Goldstein},
  {Burgess}, {Preece}, {Briggs}, {Guiriec}, {van der Horst}, {Connaughton},
  {Wilson-Hodge}, {Paciesas}, {Meegan} et~al.}}]{2012ApJS..199...19G}
\bibinfo{author}{\bibfnamefont{A.}~\bibnamefont{{Goldstein}}},
  \bibinfo{author}{\bibfnamefont{J.~M.} \bibnamefont{{Burgess}}},
  \bibinfo{author}{\bibfnamefont{R.~D.} \bibnamefont{{Preece}}},
  \bibinfo{author}{\bibfnamefont{M.~S.} \bibnamefont{{Briggs}}},
  \bibinfo{author}{\bibfnamefont{S.}~\bibnamefont{{Guiriec}}},
  \bibinfo{author}{\bibfnamefont{A.~J.} \bibnamefont{{van der Horst}}},
  \bibinfo{author}{\bibfnamefont{V.}~\bibnamefont{{Connaughton}}},
  \bibinfo{author}{\bibfnamefont{C.~A.} \bibnamefont{{Wilson-Hodge}}},
  \bibinfo{author}{\bibfnamefont{W.~S.} \bibnamefont{{Paciesas}}},
  \bibinfo{author}{\bibfnamefont{C.~A.} \bibnamefont{{Meegan}}},
  \bibnamefont{et~al.}, \bibinfo{journal}{"Astrophys. J., Suppl. Ser."}
  \textbf{\bibinfo{volume}{199}}, \bibinfo{eid}{19} (\bibinfo{year}{2012}),
  \eprint{1201.2981}.

\bibitem[{\citenamefont{{Piran}}(1999)}]{1999PhR...314..575P}
\bibinfo{author}{\bibfnamefont{T.}~\bibnamefont{{Piran}}},
  \bibinfo{journal}{Physics Reports} \textbf{\bibinfo{volume}{314}},
  \bibinfo{pages}{575} (\bibinfo{year}{1999}), \eprint{arXiv:astro-ph/9810256}.

\bibitem[{\citenamefont{{Band} et~al.}(1993)\citenamefont{{Band}, {Matteson},
  {Ford}, {Schaefer}, {Palmer}, {Teegarden}, {Cline}, {Briggs}, {Paciesas},
  {Pendleton} et~al.}}]{1993ApJ...413..281B}
\bibinfo{author}{\bibfnamefont{D.}~\bibnamefont{{Band}}},
  \bibinfo{author}{\bibfnamefont{J.}~\bibnamefont{{Matteson}}},
  \bibinfo{author}{\bibfnamefont{L.}~\bibnamefont{{Ford}}},
  \bibinfo{author}{\bibfnamefont{B.}~\bibnamefont{{Schaefer}}},
  \bibinfo{author}{\bibfnamefont{D.}~\bibnamefont{{Palmer}}},
  \bibinfo{author}{\bibfnamefont{B.}~\bibnamefont{{Teegarden}}},
  \bibinfo{author}{\bibfnamefont{T.}~\bibnamefont{{Cline}}},
  \bibinfo{author}{\bibfnamefont{M.}~\bibnamefont{{Briggs}}},
  \bibinfo{author}{\bibfnamefont{W.}~\bibnamefont{{Paciesas}}},
  \bibinfo{author}{\bibfnamefont{G.}~\bibnamefont{{Pendleton}}},
  \bibnamefont{et~al.}, \bibinfo{journal}{"Astrophys. J."}
  \textbf{\bibinfo{volume}{413}}, \bibinfo{pages}{281} (\bibinfo{year}{1993}).

\bibitem[{\citenamefont{{Racusin} et~al.}(2011)\citenamefont{{Racusin},
  {Oates}, {Schady}, {Burrows}, {de Pasquale}, {Donato}, {Gehrels}, {Koch},
  {McEnery}, {Piran} et~al.}}]{2011ApJ...738..138R}
\bibinfo{author}{\bibfnamefont{J.~L.} \bibnamefont{{Racusin}}},
  \bibinfo{author}{\bibfnamefont{S.~R.} \bibnamefont{{Oates}}},
  \bibinfo{author}{\bibfnamefont{P.}~\bibnamefont{{Schady}}},
  \bibinfo{author}{\bibfnamefont{D.~N.} \bibnamefont{{Burrows}}},
  \bibinfo{author}{\bibfnamefont{M.}~\bibnamefont{{de Pasquale}}},
  \bibinfo{author}{\bibfnamefont{D.}~\bibnamefont{{Donato}}},
  \bibinfo{author}{\bibfnamefont{N.}~\bibnamefont{{Gehrels}}},
  \bibinfo{author}{\bibfnamefont{S.}~\bibnamefont{{Koch}}},
  \bibinfo{author}{\bibfnamefont{J.}~\bibnamefont{{McEnery}}},
  \bibinfo{author}{\bibfnamefont{T.}~\bibnamefont{{Piran}}},
  \bibnamefont{et~al.}, \bibinfo{journal}{"Astrophys. J."}
  \textbf{\bibinfo{volume}{738}}, \bibinfo{eid}{138} (\bibinfo{year}{2011}),
  \eprint{1106.2469}.

\bibitem[{\citenamefont{{De Pasquale} et~al.}(2010)\citenamefont{{De Pasquale},
  {Schady}, {Kuin}, {Page}, {Curran}, {Zane}, {Oates}, {Holland}, {Breeveld},
  {Hoversten} et~al.}}]{2010ApJ...709L.146D}
\bibinfo{author}{\bibfnamefont{M.}~\bibnamefont{{De Pasquale}}},
  \bibinfo{author}{\bibfnamefont{P.}~\bibnamefont{{Schady}}},
  \bibinfo{author}{\bibfnamefont{N.~P.~M.} \bibnamefont{{Kuin}}},
  \bibinfo{author}{\bibfnamefont{M.~J.} \bibnamefont{{Page}}},
  \bibinfo{author}{\bibfnamefont{P.~A.} \bibnamefont{{Curran}}},
  \bibinfo{author}{\bibfnamefont{S.}~\bibnamefont{{Zane}}},
  \bibinfo{author}{\bibfnamefont{S.~R.} \bibnamefont{{Oates}}},
  \bibinfo{author}{\bibfnamefont{S.~T.} \bibnamefont{{Holland}}},
  \bibinfo{author}{\bibfnamefont{A.~A.} \bibnamefont{{Breeveld}}},
  \bibinfo{author}{\bibfnamefont{E.~A.} \bibnamefont{{Hoversten}}},
  \bibnamefont{et~al.}, \bibinfo{journal}{"Astrophys. J."l}
  \textbf{\bibinfo{volume}{709}}, \bibinfo{pages}{L146} (\bibinfo{year}{2010}),
  \eprint{0910.1629}.

\bibitem[{\citenamefont{{Ackermann} et~al.}(2013)\citenamefont{{Ackermann},
  {Ajello}, {Asano}, {Baldini}, {Barbiellini}, {Baring}, {Bastieri},
  {Bellazzini}, {Blandford}, {Bonamente} et~al.}}]{2013ApJ...763...71A}
\bibinfo{author}{\bibfnamefont{M.}~\bibnamefont{{Ackermann}}},
  \bibinfo{author}{\bibfnamefont{M.}~\bibnamefont{{Ajello}}},
  \bibinfo{author}{\bibfnamefont{K.}~\bibnamefont{{Asano}}},
  \bibinfo{author}{\bibfnamefont{L.}~\bibnamefont{{Baldini}}},
  \bibinfo{author}{\bibfnamefont{G.}~\bibnamefont{{Barbiellini}}},
  \bibinfo{author}{\bibfnamefont{M.~G.} \bibnamefont{{Baring}}},
  \bibinfo{author}{\bibfnamefont{D.}~\bibnamefont{{Bastieri}}},
  \bibinfo{author}{\bibfnamefont{R.}~\bibnamefont{{Bellazzini}}},
  \bibinfo{author}{\bibfnamefont{R.~D.} \bibnamefont{{Blandford}}},
  \bibinfo{author}{\bibfnamefont{E.}~\bibnamefont{{Bonamente}}},
  \bibnamefont{et~al.}, \bibinfo{journal}{"Astrophys. J."}
  \textbf{\bibinfo{volume}{763}}, \bibinfo{eid}{71} (\bibinfo{year}{2013}),
  \eprint{1212.0973}.

\bibitem[{\citenamefont{{Abdo} et~al.}(2011)\citenamefont{{Abdo}, {Ackermann},
  {Ajello}, {Baldini}, {Ballet}, {Barbiellini}, {Baring}, {Bastieri},
  {Bechtol}, {Bellazzini} et~al.}}]{2011ApJ...734L..27A}
\bibinfo{author}{\bibfnamefont{A.~A.} \bibnamefont{{Abdo}}},
  \bibinfo{author}{\bibfnamefont{M.}~\bibnamefont{{Ackermann}}},
  \bibinfo{author}{\bibfnamefont{M.}~\bibnamefont{{Ajello}}},
  \bibinfo{author}{\bibfnamefont{L.}~\bibnamefont{{Baldini}}},
  \bibinfo{author}{\bibfnamefont{J.}~\bibnamefont{{Ballet}}},
  \bibinfo{author}{\bibfnamefont{G.}~\bibnamefont{{Barbiellini}}},
  \bibinfo{author}{\bibfnamefont{M.~G.} \bibnamefont{{Baring}}},
  \bibinfo{author}{\bibfnamefont{D.}~\bibnamefont{{Bastieri}}},
  \bibinfo{author}{\bibfnamefont{K.}~\bibnamefont{{Bechtol}}},
  \bibinfo{author}{\bibfnamefont{R.}~\bibnamefont{{Bellazzini}}},
  \bibnamefont{et~al.}, \bibinfo{journal}{"Astrophys. J."l}
  \textbf{\bibinfo{volume}{734}}, \bibinfo{eid}{L27} (\bibinfo{year}{2011}),
  \eprint{1104.5496}.

\bibitem[{\citenamefont{{Abeysekara} et~al.}(2012)\citenamefont{{Abeysekara},
  {Aguilar}, {Aguilar}, {Alfaro}, {Almaraz}, {{\'A}lvarez},
  {{\'A}lvarez-Romero}, {{\'A}lvarez}, {Arceo}, {Arteaga-Vel{\'a}zquez}
  et~al.}}]{2012APh....35..641A}
\bibinfo{author}{\bibfnamefont{A.~U.} \bibnamefont{{Abeysekara}}},
  \bibinfo{author}{\bibfnamefont{J.~A.} \bibnamefont{{Aguilar}}},
  \bibinfo{author}{\bibfnamefont{S.}~\bibnamefont{{Aguilar}}},
  \bibinfo{author}{\bibfnamefont{R.}~\bibnamefont{{Alfaro}}},
  \bibinfo{author}{\bibfnamefont{E.}~\bibnamefont{{Almaraz}}},
  \bibinfo{author}{\bibfnamefont{C.}~\bibnamefont{{{\'A}lvarez}}},
  \bibinfo{author}{\bibfnamefont{J.~d.~D.} \bibnamefont{{{\'A}lvarez-Romero}}},
  \bibinfo{author}{\bibfnamefont{M.}~\bibnamefont{{{\'A}lvarez}}},
  \bibinfo{author}{\bibfnamefont{R.}~\bibnamefont{{Arceo}}},
  \bibinfo{author}{\bibfnamefont{J.~C.} \bibnamefont{{Arteaga-Vel{\'a}zquez}}},
  \bibnamefont{et~al.}, \bibinfo{journal}{Astroparticle Physics}
  \textbf{\bibinfo{volume}{35}}, \bibinfo{pages}{641} (\bibinfo{year}{2012}),
  \eprint{1108.6034}.

\bibitem[{\citenamefont{{Kakuwa} et~al.}(2012)\citenamefont{{Kakuwa}, {Murase},
  {Toma}, {Inoue}, {Yamazaki}, and {Ioka}}}]{2012MNRAS.425..514K}
\bibinfo{author}{\bibfnamefont{J.}~\bibnamefont{{Kakuwa}}},
  \bibinfo{author}{\bibfnamefont{K.}~\bibnamefont{{Murase}}},
  \bibinfo{author}{\bibfnamefont{K.}~\bibnamefont{{Toma}}},
  \bibinfo{author}{\bibfnamefont{S.}~\bibnamefont{{Inoue}}},
  \bibinfo{author}{\bibfnamefont{R.}~\bibnamefont{{Yamazaki}}},
  \bibnamefont{and} \bibinfo{author}{\bibfnamefont{K.}~\bibnamefont{{Ioka}}},
  \bibinfo{journal}{"Mon. Not. R. Astron. Soc."}
  \textbf{\bibinfo{volume}{425}}, \bibinfo{pages}{514} (\bibinfo{year}{2012}),
  \eprint{1112.5940}.

\bibitem[{\citenamefont{{Abadie} et~al.}(2012)\citenamefont{{Abadie}, {Abbott},
  {Abbott}, {Abbott}, {Abernathy}, {Accadia}, {Acernese}, {Adams}, {Adhikari},
  {Affeldt} et~al.}}]{2012ApJ...760...12A}
\bibinfo{author}{\bibfnamefont{J.}~\bibnamefont{{Abadie}}},
  \bibinfo{author}{\bibfnamefont{B.~P.} \bibnamefont{{Abbott}}},
  \bibinfo{author}{\bibfnamefont{R.}~\bibnamefont{{Abbott}}},
  \bibinfo{author}{\bibfnamefont{T.~D.} \bibnamefont{{Abbott}}},
  \bibinfo{author}{\bibfnamefont{M.}~\bibnamefont{{Abernathy}}},
  \bibinfo{author}{\bibfnamefont{T.}~\bibnamefont{{Accadia}}},
  \bibinfo{author}{\bibfnamefont{F.}~\bibnamefont{{Acernese}}},
  \bibinfo{author}{\bibfnamefont{C.}~\bibnamefont{{Adams}}},
  \bibinfo{author}{\bibfnamefont{R.~X.} \bibnamefont{{Adhikari}}},
  \bibinfo{author}{\bibfnamefont{C.}~\bibnamefont{{Affeldt}}},
  \bibnamefont{et~al.}, \bibinfo{journal}{"Astrophys. J."}
  \textbf{\bibinfo{volume}{760}}, \bibinfo{eid}{12} (\bibinfo{year}{2012}),
  \eprint{1205.2216}.

\bibitem[{\citenamefont{{LIGO Scientific Collaboration}
  et~al.}(2013)\citenamefont{{LIGO Scientific Collaboration}, {Virgo
  Collaboration}, {Aasi}, {Abadie}, {Abbott}, {Abbott}, {Abbott}, {Abernathy},
  {Accadia}, {Acernese} et~al.}}]{2013arXiv1304.0670L}
\bibinfo{author}{\bibnamefont{{LIGO Scientific Collaboration}}},
  \bibinfo{author}{\bibnamefont{{Virgo Collaboration}}},
  \bibinfo{author}{\bibfnamefont{J.}~\bibnamefont{{Aasi}}},
  \bibinfo{author}{\bibfnamefont{J.}~\bibnamefont{{Abadie}}},
  \bibinfo{author}{\bibfnamefont{B.~P.} \bibnamefont{{Abbott}}},
  \bibinfo{author}{\bibfnamefont{R.}~\bibnamefont{{Abbott}}},
  \bibinfo{author}{\bibfnamefont{T.~D.} \bibnamefont{{Abbott}}},
  \bibinfo{author}{\bibfnamefont{M.}~\bibnamefont{{Abernathy}}},
  \bibinfo{author}{\bibfnamefont{T.}~\bibnamefont{{Accadia}}},
  \bibinfo{author}{\bibfnamefont{F.}~\bibnamefont{{Acernese}}},
  \bibnamefont{et~al.}, \bibinfo{journal}{ArXiv e-prints}
  (\bibinfo{year}{2013}), \eprint{1304.0670}.

\end{thebibliography}
\end{document}